\documentclass[12pt,notitlepage,a4paper]{article}
\pdfoutput=1

\usepackage{epsfig}
\usepackage{color,graphicx}
\usepackage{cite}
\usepackage{amssymb,amsmath}

\newcommand{\be}{\begin{equation}}
\newcommand{\ee}{\end{equation}}
\newcommand{\bea}{\begin{eqnarray}}
\newcommand{\eea}{\end{eqnarray}}

\begin{document}

\begin{flushright}
CCTP-2011-19 
\end{flushright}

\begin{center}  

\vskip 2cm 

\centerline{\Large {\bf Fluctuations of a holographic quantum Hall fluid}}
\vskip 1cm

\renewcommand{\thefootnote}{\fnsymbol{footnote}}

\centerline{Niko Jokela,${}^{1,2}$\footnote{najokela@physics.technion.ac.il}
Matti J\"arvinen,${}^3$\footnote{mjarvine@physics.uoc.gr} 
and Matthew Lippert${}^3$\footnote{mlippert@physics.uoc.gr}}

\vskip .5cm
${}^1${\small \sl Department of Physics} \\
{\small \sl Technion, Haifa 32000, Israel} 

\vskip 0.2cm
${}^2${\small \sl Department of Mathematics and Physics} \\
{\small \sl University of Haifa at Oranim, Tivon 36006, Israel} 

\vskip 0.2cm
${}^3${\small \sl Crete Center for Theoretical Physics} \\
{\small \sl Department of Physics} \\
{\small \sl University of Crete,  71003 Heraklion, Greece}

\end{center}

\vskip 0.3 cm

\setcounter{footnote}{0}
\renewcommand{\thefootnote}{\arabic{footnote}}

\begin{abstract}
\noindent  
We analyze the neutral spectrum of the holographic quantum Hall fluid described by the D2-D8' model.  As expected for a quantum Hall state, we find the system to be stable and gapped and that typically the lowest excitation mode is a magneto-roton.  In addition, we find magneto-rotons in higher modes as well.  We show that these magneto-rotons are direct consequences of level crossings between vector and scalar modes.
\end{abstract}

\newpage

\tableofcontents

\section{Introduction}
\label{sec:intro}

The quantum Hall (QH) effect  describes the behavior of electrons in a two-dimensional conductor in a transverse magnetic field.\footnote{For a review of the quantum Hall effect, see \cite{girvin}.}  When the filling fraction, defined as the ratio of charge density to the magnetic field in units of 
$\frac{e}{2\pi \hbar}$ and denoted by $\nu$, takes integer (integer QH effect) and certain rational (fractional QH effect) values, the electron fluid enters a QH phase with many striking properties.  Most characteristic of these, of course, is the response to an applied electric field; the conductivity in the direction of the field vanishes, while the conductivity in the orthogonal direction, or Hall conductivity, is exactly $\frac{e^2}{2\pi \hbar} \nu$.

The fluctuations of QH fluids exhibit many interesting properties which have been observed experimentally.  In the QH state, a mass gap is dynamically generated for both the charged and neutral excitations.  The charged fluctuations are quasiparticles, which in the fractional QH effect are fractionally charged and which obey anyon statistics.  

The lowest neutral excitations of quantum Hall fluids are magneto-rotons. These are collective excitations whose minimum energy is at zero group velocity but nonzero momentum.  Rotons were first observed in superfluids, as local minima in the phonon dispersion at large momentum.  The magneto-rotons in quantum Hall fluids occur as global minima of the energy at nonzero momentum and have been detected in many experiments \cite{light, ballistic, recent}.  There are several theoretical many-body and field theoretic treatments \cite{Girvin:1986zz, spj, cs}.

The important features of the QH effect are universal, in that they do not depend significantly on the details of the experiment, such as, for example, the crystal structure of the material. This kind of generic phenomenon makes a promising candidate for holographic modeling.

Our program is to investigate a class of top-down holographic models of strongly-coupled (2+1)-dimensional fermions exhibiting quantum Hall behavior.\footnote{The QH effect has also been studied using several other types of holographic models \cite{otherqh}.}  We have so far studied two similar brane constructions, the D3-D7' model of the fractional QH effect  \cite{Bergman:2010gm} and the D2-D8' model of the integer QH effect \cite{Jokela:2011eb}.  In both of these models, the QH state corresponds holographically to an embedding of the flavor brane which does not enter the black hole horizon.  This feature of the embedding leads directly to a dynamically-induced mass gap and the expected conductivities for a QH fluid.

The neutral fluctuations of the holographic QH state of the D3-D7' model were studied recently in \cite{Jokela:2010nu}.\footnote{An analogous fluctuation analysis of the ungapped, metallic state of the D3-D7' system was conducted in \cite{Bergman:2011rf}.} That investigation was limited to zero temperature.  As expected, the spectrum was found to be gapped, indicating that the fermion fluid is incompressible.  A tachyon was found in the branch of solutions previously identified as thermodynamically unstable \cite{Bergman:2010gm}, but otherwise the QH state was stable.  The lowest mode was shown to become rotonic in the vicinity of the crossing of the lowest scalar mode and the lowest vector mode. 

In this paper, we perform a similar fluctuation analysis of the QH state of the D2-D8' system.  We investigate the neutral bosonic spectrum and improve upon the D3-D7' study by considering nonzero temperature.  As expected, we again find a mass gap, with a tachyon only in the thermodynamically unstable solution.  We also show that, unlike in the D3-D7' model, the speed of sound is independent of mode number and weakly dependent on charge density.  We find that, over a wide range of parameters, the lowest mode is a magneto-roton, and in addition, magneto-rotons appear in higher modes as well.  We provide an analytical argument showing how these magneto-rotons are in fact direct consequences of level crossings between scalar and vector modes.

In the next section, Section \ref{sec:review}, we review the D2-D8' model, its action and solutions.  We then set up the analysis of the fluctuations and describe our numerical methods in Section \ref{sec:fluct}.  Section \ref{sec:numerics} contains our results, the spectrum and dispersion of the modes.  We investigate the relationship between the observed level crossings and the rotonic dispersions in Section \ref{sec:levelcrossingandrotons}.  Finally, we conclude in Section \ref{sec:discussion} with a discussion and open questions.

\section{Review of the D2-D8' model}
\label{sec:review}
The D2-D8' system \cite{Jokela:2011eb} consists of a probe D8-brane in the near-horizon background of $N$ thermal D2-branes, such that $\# ND=6$ in the flat-space limit. We work in the limit where $N \gg 1$ and $g_s N \gg 1$. Supersymmetry is completely broken, and internal flux is needed for stability.  At weak coupling, the low-energy spectrum of bifundamental strings in a $\# ND=6$ system contains only charged fermions and no charged bosons; as a result, the holographic field theory dual is (2+1)-dimensional SYM coupled to $N$ species of charged fermions.  Similar constructions have been used, for example, in the D3-D7 \cite{Myers:2008me} and D3-D7' \cite{Bergman:2010gm} models.  

\subsection{Set-up}
The ten-dimensional, near-horizon D2-brane metric is
\be
L^{-2} ds^2_{10} = u^\frac{5}{2}\left(-f(u)dt^2+dx^2+dy^2\right)+u^{-\frac{5}{2}}\left(\frac{du^2}{f(u)}+ u^2 d\Omega_{S^6}^2\right) \ ,
\ee
where $L^5 = 6\pi^2 g_s N l_s^5$ and the thermal factor $f(u)=1-\left(\frac{u_T}{u}\right)^5$.  There is a black hole horizon at $u = u_T$, and the corresponding Hawking temperature is $T= \frac{5}{4\pi L} u_T^{3/2}$.  The metric on the $S^6$ is given by
\be
d\Omega_{S^6}^2 = d\psi^2 + \sin^2\psi \left(d\theta_1^2 + \sin^2\theta_1 d\phi_1^2\right) + \cos^2\psi \left(d\xi^2 + \sin^2\xi d\theta_2^2+ \sin^2\xi \sin^2\theta_2 d\phi_2^2\right) \ ,
\ee
where $\psi$ ranges from 0 to $\pi/2$, $\xi$, $\theta_1$, and $\theta_2$ range from 0 to $\pi$, and $\phi_1$ and $\phi_2$ range from 0 to $2\pi$. The background dilaton and RR 5-form potential are
\bea
e^\phi &=& g_s\left(\frac{L}{U}\right)^\frac{5}{4} \\
C^{(5)} &=& c(\psi) L^5  d\Omega_{S^2} \wedge d\Omega_{S^3} \ ,
\eea
where $c(\psi) = \frac{5}{8}\left(\sin\psi - \frac{1}{6}\sin(3\psi)- \frac{1}{10}\sin(5\psi)\right)$.

We will work in  dimensionless (lowercase) coordinates, which are related to the physical dimensionful (uppercase) 
coordinates as: $x^\mu = X^\mu/L$, $u = U/L$.  In general, lowercase Latin letters will denote dimensionless quantities, 
and uppercase letters will denote the corresponding physical quantities.

The probe D8-brane fills the spacetime directions $t$, $x$, $y$, and the radial direction $u$ and wraps an $S^2 \times S^3$ fibered over an interval in the internal $S^6$.  The embedding in the $\psi$-direction is a function of $u$.  The induced metric on the D8-brane is then
\bea
L^{-2}ds^2_{D8} &=& u^\frac{5}{2}\left(-f dt^2+dx^2+dy^2\right)+u^{-\frac{5}{2}}\left(\frac{1}{f} +u^2\psi'^2\right)du^2 \nonumber\\
&&+ u^{-\frac{1}{2}}\sin^2\psi \ d\Omega_2^2 + u^{-\frac{1}{2}}\cos^2\psi \ d\Omega_{3}^2 \ ,
\eea
where $' \equiv \partial_u$.

We would like to consider a boundary system with nonzero charge density in a background magnetic field. In the bulk, this is accomplished by turning on components of the worldvolume gauge field $A_\mu$; specifically, a constant field strength in the spatial directions corresponds to a transverse background magnetic field, and nonzero charge density is dual to a radially varying electric field:
\bea
2\pi\alpha' F_{xy} &=& h \\
2\pi\alpha' F_{u0} &=& a_0'(u) \ .
\eea
In addition, we include a magnetic field wrapping the internal $S^2$ in order to stabilize the embedding:
\be
2\pi\alpha'F_{\theta_1\phi_1} = L^2 b \sin \phi_1 \ .
\ee
Furthermore, we choose to work in radial gauge, where $A_u = 0$.

The DBI action for the D8-brane is then
\bea
 S_{DBI} &=&  - \mu_8 \int d^9x \, e^{-\phi} \sqrt{-\det(g_{\mu\nu} + 2\pi\alpha' F_{\mu\nu})}  \\
&=&  - \mathcal N \int du  \, u^\frac{5}{2} \cos^3\psi \sqrt{\left(1+f u^2\psi'^2- a_0'^2\right )\left(b^2u + \sin^4\psi\right)\left(1+\frac{h^2}{u^5}\right)} \quad ,
\label{DBIaction}
\eea
where $\mathcal N = 8 \pi^3 T_8 v_3 L^9$ and $v_3$ is the dimensionless volume of spacetime.  As a result of the charge density and the magnetic field, the Chern-Simons action has a nonzero term:
\be\label{CSaction}
S_{CS} = -\frac{T_8}{2}(2\pi\alpha')^2\int C^{(5)}\wedge F\wedge F = \mathcal N \int du \, c(\psi) ha_0' \ .
\ee

\subsection{Background solution}
The D8-brane action (\ref{DBIaction}) and (\ref{CSaction}) is independent of $a_0$, so the equation of motion for $a_0$ can be integrated once, giving
\be
\frac{g}{f}\left(1 + \frac{h^2}{u^5}\right) a_0' = d-hc(\psi) \equiv \tilde d(u) \ ,
\ee
where we have defined
\be
\label{ga0}
g(u) = \frac{f u^{5/2}\cos^3\psi \sqrt{b^2u + \sin^4\psi}}{\sqrt{\left(1+f u^2\psi'^2-a'^2_0\right)\left(1 + \frac{h^2}{u^5}\right)}} 
\ee
and where $d$ is the constant of integration and $\tilde d(u)$ is the radial displacement field.  In terms of the boundary theory, $d$ is the total charge density, while $\tilde d(u)$ is the charge due to sources located in the bulk at radial positions below $u$.

We can solve for $a_0'$ in terms of $\tilde d$, obtaining
\be
a_0' = \tilde d \sqrt{\frac{1+fu^2\psi'^2}{\tilde d^2 + u^5\cos^6\psi \left(b^2u + \sin^4\psi\right)\left(1 + \frac{h^2}{u^5}\right) }} \quad .
\ee
We then rewrite (\ref{ga0}) as
\be
\label{g}
g = \frac{f}{1+\frac{h^2}{u^5}} \sqrt{\frac{\tilde d^2 + u^5\cos^6\psi\left(1 + \frac{h^2}{u^5}\right)\left(b^2u + \sin^4\psi\right)}{1+f u^2\psi'^2}} \quad .
\ee

The $\psi$ equation of motion, written in terms of $\tilde d$, is
\bea
\label{psieom}
\partial_u\left(u^2 g \left(1 + \frac{h^2}{u^5}\right) \psi'\right) &=& \frac{f u^5}{g} \cos^5\psi\sin\psi\left(2\cos^2\psi\sin^2\psi - 3\left(b^2u + \sin^4\psi\right)\right) \nonumber \\ &&- 5 \frac{f}{g}\frac{h \tilde d}{1+\frac{h^2}{u^5}}\cos^3\psi \ .
\eea
We can solve (\ref{psieom}) at large $u$ and find that $\psi \to 0$ in the UV.  We identify the coefficient of the leading term in the asymptotic expansion of $\psi$ as the fermion mass and the coefficient of the subleading term as the chiral condensate: 
\be
\psi = \frac{m}{u} - \frac{c}{u^3} + \mathcal O(u^{-4}) \ .
\ee

There are two classes of D8-brane embeddings with different IR behavior: black hole (BH) embeddings, which enter the horizon at $u_T$, and Minkowski (MN) embeddings, which smoothly cap off at some $u_0 > u_T$ as the wrapped $S^3$ shrinks to zero size.  As argued in \cite{Jokela:2011eb}, the MN embeddings holographically reproduce the properties of a quantum Hall fluid, while the BH solutions exhibit metallic behavior.  In this paper, we will focus on fluctuations of the MN embedding, reserving the study of the quasinormal modes of the BH solutions for a future work.  

The QH features of the MN solution are direct consequences of the boundary conditions at the tip, $\psi(u_0) = \pi/2$.  We found in \cite{Jokela:2011eb} that regularity of the gauge fields requires that the ratio of the charge density $d$ and the magnetic field $h$ be fixed:
\be
\frac{d}{h} = c(\pi/2) = \frac{2}{3} \ .
\ee
The filling fraction per fermion species $\nu$ is defined by the ratio\footnote{The filling fraction $\nu$ is given in units of the Dirac flux quantum $\frac{e}{2\pi\hbar}$, which is just $\frac{1}{2\pi}$ in units where $e=\hbar =1$.} of the physical charge density $D = \frac{(2\pi\alpha')\mathcal N}{L^4 v_3} d$ to the physical magnetic field $H = \frac{h}{2\pi\alpha'}$, divided by the number of species $N$.  We find that
\be
\nu = \frac{2\pi}{N} \frac{D}{H} = \frac{3}{2} \frac{d}{h} = 1 \ .
\ee
The MN embedding is therefore dual to an integer QH fluid.  

We found evidence in \cite{Jokela:2011eb} to support this identification.  By avoiding the horizon, the MN embedding generates a dynamically-induced mass gap for charged excitations, given by $u_0$.  In Section \ref{sec:numerics}, we will find a mass gap for neutral fluctuations as well.  In addition, as expected in a QH state, the Hall conductivity is given by the filling fraction, $\sigma_{xy} = \frac{\nu}{2\pi}$, while the longitudinal conductivity vanishes, $\sigma_{xx}=0$.

\subsection{Isotropic coordinates} 
To investigate the fluctuations of the D8-brane, it is much more convenient to work in Cartesian coordinates rather than the polar coordinates $(u, \psi)$.  This is simple to accomplish at zero temperature; however, at nonzero temperature, because of the thermal factor $f$, we first need to convert to a new, isotropic radial coordinate $r$ where
\be
r^{5/2} = \left(\frac{u}{u_T}\right)^{5/2} + \sqrt{\left(\frac{u}{u_T}\right)^{5} - 1} \ .
\ee
The horizon is located at $r=1$, and the boundary is at $r =\infty$. For convenience, we define
\be
\beta = \frac{1-r^{-5}}{2} \qquad , \qquad \tilde\beta = \frac{1+r^{-5}}{2} \ .
\ee
This allows the thermal factor to be written as 
\be
f =\left(\frac{\beta}{\tilde\beta}\right)^{2} \ ,
\ee
and the ten-dimensional metric takes the form
\be
L^{-2}ds_{10}^2 = (u_T r)^{5/2}\tilde\beta\left(-\frac{\beta^2}{\tilde\beta^2}dt^2 + dx^2 + dy^2\right)+ \frac{1}{u_T^{1/2}r^{5/2}\tilde\beta^{1/5}} \left(dr^2 + r^2d\Omega_6^2 \right) \ .
\ee

We can now introduce Cartesian coordinates on the $(r,\psi)$-plane:
\bea
\rho &=& r \sin\psi \\
R &=& r \cos\psi \ ,
\eea
where $\rho$ and $R$ give the sizes of the $S^2$ and $S^3$, respectively.  We parametrize the D8-brane embedding as $\rho(R)$, and the radial gauge field is $2\pi\alpha'F_{R0} = a_0'(R)$, where the prime now denotes partial derivatives with respect to $R$.  The induced metric on the D8-brane is then
\bea
L^{-2}ds_{D8}^2 &=& (u_T r)^{5/2}\tilde\beta\left(-\frac{\beta^2}{\tilde\beta^2}dt^2 + dx^2 + dy^2\right) \nonumber\\
& & + \frac{1}{u_T^{1/2}r^{5/2}\tilde\beta^{1/5}} \left((1+\rho'^2)dR^2 + \rho^2 d\Omega_2^2 + R^2 d\Omega_3^2  \right) \ .
\eea
The D8-brane DBI action in these coordinates is
\be
\label{DBIactionisotropic}
S_{DBI} = -\mathcal N\! \int\! dR \ u_T^4 R^3 \beta\tilde\beta^{3/5}\! \sqrt{\!\left(1+\rho'^2-\frac{\tilde\beta^{6/5}}{u_T^2\beta^2}a_0'^2\right)\!\!\left(1+\frac{h^2}{u_T^5 r^5 \tilde\beta^2}\right)\!\!\left(\frac{\rho^4}{u_T r^5\tilde\beta^{2/5}} + b^2\right)}
\ee
and the Chern-Simons action is
\be
\label{CSactionisotropic}
S_{CS} = \mathcal N \int dR \ c(\rho)ha_0' \ ,
\ee
where $c(\rho) = \frac{5\rho^3}{3r^3} - \frac{\rho^5}{r^5}$.

\begin{figure}[ht]
\center
\includegraphics[width=0.7\textwidth]{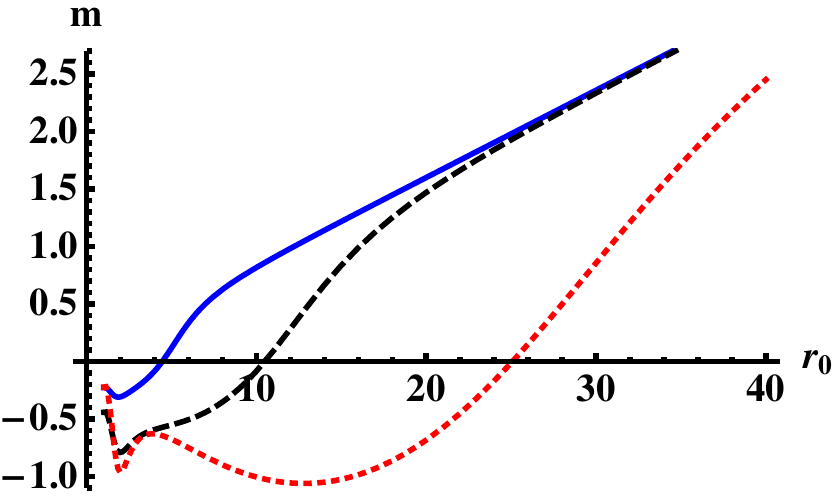}
\caption{The fermion mass $m$ plotted as a function of $r_0$ for $u_T = 0.1$ and $b=1$.  The three curves correspond to different values of $d$: $d=0.1$ solid blue, $d=1$ dashed black, and $d=10$ dotted red.  Note that, embeddings that seemingly yield a negative mass are actually unstable due to self-intersections \cite{Jokela:2011eb}.}
\label{fig:mvsr0}
\end{figure}

The background MN embedding caps off smoothly at $r=r_0$ as the $S^3$ collapses, which implies the IR boundary condition $\rho(R=0) = r_0$.  We still choose a radial gauge, which is now $A_R = 0$.  Solving the equations of motion derived from (\ref{DBIactionisotropic}) and (\ref{CSactionisotropic}), we find solutions for the embedding $\bar\rho(R)$ and radial gauge field $\bar a_0(R)$.  The fermion mass is related to the asymptotic embedding by
\be
\bar\rho(\infty) = 2^{2/5} \frac{m}{u_T} \ .
\ee
The mass as a function of $r_0$ for several values of $d$ is shown in Fig.~\ref{fig:mvsr0}.

\section{Fluctuations}
\label{sec:fluct}

\subsection{Expansion}
We now allow for fluctuations of the following worldvolume fields:
\bea
 \rho        & = & \bar\rho(R)+\epsilon \, \delta \rho(t,x,y,R) \\
 a_0         & = & \bar a_0(R)+\epsilon \, \delta a_0(t,x,y,R) \\
 a_x         & = & \epsilon \, \delta a_x(t,x,y,R) \\
 a_y         & = & h  x +\epsilon \, \delta a_y(t,x,y,R) \\
 a_R        & = & \epsilon \, \delta a_{R}(t,x,y,R) \ ,
\eea
with $\epsilon$ a small parameter and where $a_\mu = \frac{1}{2\pi\alpha'} A_\mu$.  We do not consider fluctuations in the internal spheres or dependence on the internal $S^2$ or $S^3$ coordinates.  
The determinant $\det (G+2\pi\alpha' F)$ in the DBI action is now
\bea \label{DBIfluctuations}
 \left| \begin{array}{cccc}
-(u_T r)^{5/2}\frac{\beta^2}{\tilde\beta} & \epsilon(\frac{\partial\delta a_0}{\partial x} - \delta\dot a_x) & \epsilon(\frac{\partial\delta a_0}{\partial y} -\delta\dot a_y) & \bar a_0' + \epsilon(\delta a_0' - \delta\dot a_R)\\
 -\epsilon(\frac{\partial\delta a_0}{\partial x} - \delta\dot a_x) & (u_T r)^{5/2}\tilde\beta & -h+\epsilon \partial_{[y}\delta a_{x]} & \epsilon(\delta a_x'-\frac{\partial\delta a_R}{\partial x})\\
- \epsilon(\frac{\partial\delta a_0}{\partial y} -\delta\dot a_y)   & h + \epsilon \partial_{[x}\delta a_{y]}  & (u_T r)^{5/2}\tilde\beta  &  \epsilon(\delta a_y'-\frac{\partial\delta a_R}{\partial y}) \\
- \bar a_0' - \epsilon(\delta a_0' - \delta\dot a_R) & -\epsilon( \delta a_x'-\frac{\partial\delta a_R}{\partial x}) & - \epsilon( \delta a_y'-\frac{\partial\delta a_R}{\partial y}) & \frac{1+\rho'^2}{u_T^{1/2}r^{5/2}\tilde\beta^{1/5}}  \\
\end{array} \right| &&\\ 
\times \frac{R^6}{u_T^{3/2}r^{15/2}\tilde\beta^{3/5}} \left(\frac{\rho^4}{u_T r^5 \tilde\beta^{2/5}} + b^2\right)\sin^2\theta_1\sin^2\xi \sin^4\theta_2
\ ,\nonumber
\eea
where dots denote derivatives with respect to time and primes are derivatives with respect to $R$.  The Chern-Simons action becomes
\bea \label{CSfluctuations}
S_{CS} &=& \mathcal N \int dR \ c(\bar\rho +\epsilon\delta\rho)
\epsilon^2 \left( \left(\frac{\bar a_0'}{\epsilon} + \delta a_0' - \delta\dot a_R \right)\left(\frac{h}{\epsilon} + \partial_{[x}\delta a_{y]} \right) \right. \nonumber\\
&&- \left(\delta\dot a_x -\frac{\partial\delta a_0}{\partial x}\right)\left(\frac{\partial\delta a_R}{\partial y} - \delta a_y'\right)
\left. + \left(\delta\dot a_y -\frac{\partial\delta a_0}{\partial y}\right)\left(\frac{\partial\delta a_R}{\partial x} - \delta a_x'\right) \right) \ .
\eea 

The on-shell Lagrangian can then be expanded to second order in $\epsilon$:
\be
 \mathcal L = \mathcal L_0 + \epsilon\mathcal L_1 + \epsilon^2\mathcal L_2 + \ldots \ .
\ee
Here $\mathcal L_0$ corresponds to the background Lagrangian, and $\mathcal L_1$ vanishes by the background equations of motion.  We derive the equations of motion for $\delta \rho$, $\delta a_0$, $\delta a_x$,  and $\delta a_y$ from $\mathcal L_2$. However, since they are very long expressions, we will not reproduce them here.  In addition, since we are working in the radial gauge $a_R = 0$, the equation of motion for $\delta a_R$ becomes a constraint to be imposed on the fluctuations.

We make the following wavelike ansatz for the fluctuations:
\bea
  \delta \rho(t,x,R) & = & \delta \tilde\rho(R) e^{-i\omega t+ik x} \\
  \delta a_\mu(t,x,R)  & = & \delta \tilde a_\mu(R) e^{-i\omega t+ik x} \ ,
\eea
where, because of the rotational symmetry in the $(x,y)$-plane, we can choose the momentum to be in the $x$-direction without loss of generality.

We must be careful to ensure we consider physical fluctuations rather than gauge artifacts.  To ensure this, we work with the fluctuation of the longitudinal electric field:
\be
\delta \tilde e_x = k \delta\tilde a_0 + \omega \delta \tilde a_x 
\ee
which is gauge invariant.  As long as we start with initial conditions in radial gauge at some $R$ which satisfy the constraint given by the equation of motion for $\delta a_R$, the constraint and gauge condition will be satisfied by the solution at all $R$.

\subsection{Numerical methodology}
Our goal is to solve the equations of motion for the fluctuations and find the normalizable modes.  The condition for normalizability is that the fluctuations vanish at the boundary:
\be
\delta \tilde e_x(\infty) = \delta \tilde a_y(\infty) = \delta \tilde\rho(\infty) = 0 \ .
\ee
Furthermore, the embedding of the D8-brane must be smooth at the tip, which implies
\be
\delta \tilde \rho'(0) = 0 \ ,
\ee
and the gauge fields must also be nonsingular, giving
\be
 \delta \tilde e_x'(0) = \delta \tilde a_y'(0) = 0 \ .
\ee

In practice, to numerically solve the coupled equations for the fluctuations at fixed values of $\omega$ and $k$, we vary the initial conditions at the tip and shoot out to the boundary to look for solutions which are normalizable.  We must solve three coupled, linear, second-order differential equations for three independent functions.

An efficient technique for finding such solutions is the determinant method \cite{Amado:2009ts, Kaminski:2009dh}.  Rather than attempting to scan, for each $\omega$ and $k$, the full three-dimensional space of initial conditions, we choose a basis of linearly-independent initial conditions,
\be
\label{basis}
(\delta \tilde\rho(0), \delta \tilde e_x(0), \delta \tilde a_y(0)) = \left\{(1,0,0), (0,1,1), (0,1,-1)\right\} \, 
\ee
for which we numerically solve the equations of motion.  We then compute the following determinant of boundary values:
\be
 \det\left( \begin{array}{ccc}
  \delta \tilde \rho^I & \delta\tilde \rho^{II} & \delta\tilde \rho^{III} \\
 \delta\tilde e_x^I & \delta\tilde e_x^{II} & \delta\tilde e_x^{III} \\
 \delta\tilde a_y^I & \delta\tilde a_y^{II} & \delta\tilde a_y^{III} 
\end{array} \right)\Bigg|_{R\to \infty} \ ,
\ee
where the superscripts refer to the basis elements of (\ref{basis}).  If the determinant vanishes, there is at least one linear combination of the basis vectors for which $\delta\tilde\rho(\infty)$, $\delta\tilde e_x(\infty)$, 
and $\delta\tilde a_y(\infty)$ simultaneously vanish, signaling the existence of a normalizable mode.  
We repeat this procedure for various values of $\omega$ and $k$; for $k=0$, a scan in $\omega$ yields the spectrum, and allowing $k > 0$ enables us to compute the dispersions.

For the figures presented in Section \ref{sec:numerics}, we have chosen particular values of the parameters $u_T$, $m$, $d$, $h$, and $b$ for illustrative purposes.   As we will be studying exclusively MN embeddings, we will work with the parameter $d$ and always set $h = \frac{3}{2} d$. The scaling $u \to \Lambda u$, $u_T \to \Lambda u_T$, $m \to \Lambda m$, $d \to \Lambda^{5/2} d$, $h \to \Lambda^{5/2} h$, and $b \to \Lambda^{-1/2} b$ is a symmetry of the equations of motion, and so there is, in fact, only a three-dimensional parameter space of MN solutions.  Many of the qualitative results we found appear to be independent of these choices.  However, as an exhaustive numerical search is essentially impossible, we can not conclusively rule out the possibility of different behavior in particular corners of the parameter space.

\begin{figure}[ht]
\center
\includegraphics[width=0.7\textwidth]{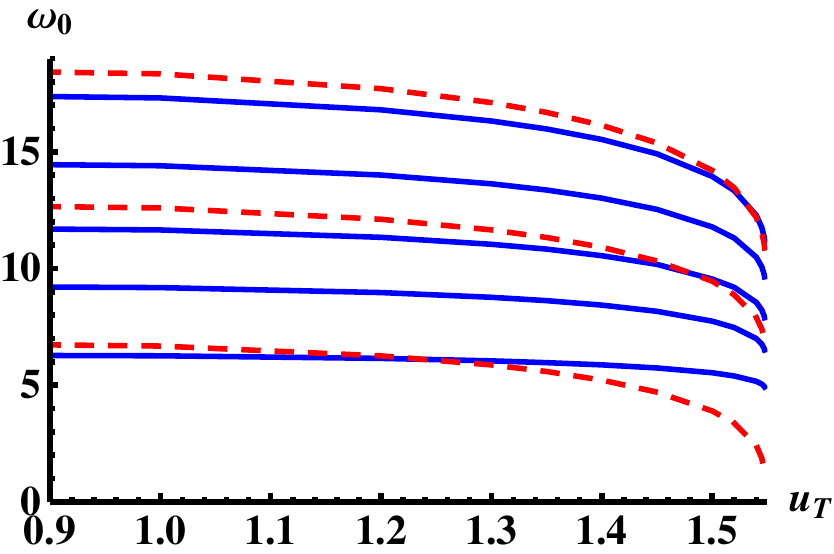}
\caption{The mass $\omega_0$ of the eight lowest modes plotted against $u_T$, for $d=10$, $b=1$ and $m=0.1$.  Blue solid curves are vector modes and red dashed curves are scalars.  There are three mode crossings: modes one and two cross at $u_T \approx 1.25$, modes four and five cross at $u_T \approx 1.50$, and modes seven and eight cross at $u_T \approx 1.54$. The maximum temperature for the MN embedding is $u_T \approx 1.548$.}
\label{fig:spectrum}
\end{figure}

\begin{figure}[ht]
\center
\includegraphics[width=0.46\textwidth]{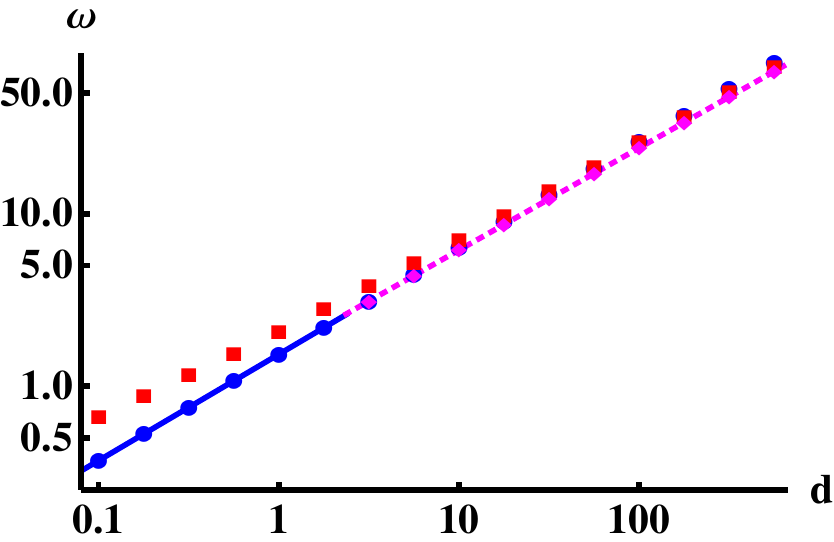}
\includegraphics[width=0.46\textwidth]{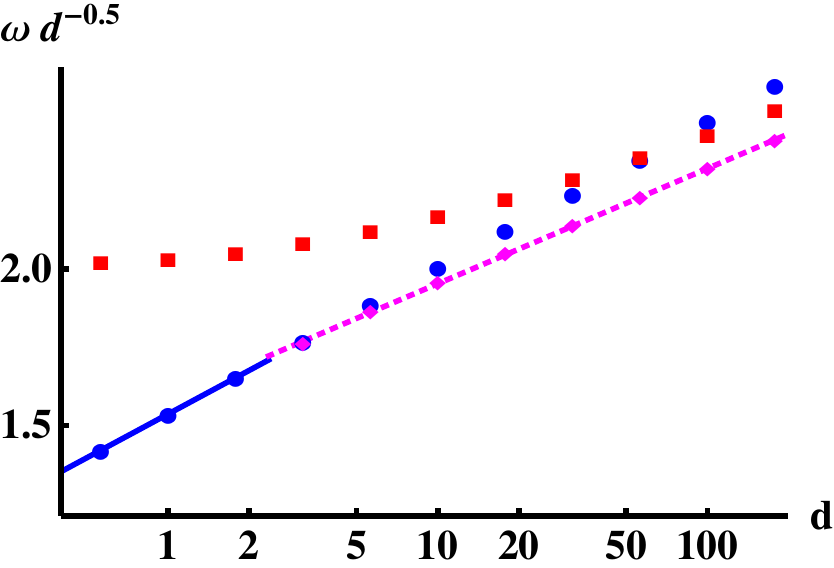}
\caption{The mass gap as a function of $d$ for $u_T = 0.1$, $b=1$, and $m=0.1$.  The blue circles show $\omega_0$ for the lowest vector mode, and the red squares for lowest scalar mode. The two levels cross at $d \approx 70$.  The magenta diamonds, which lie slightly below the $\omega_0$ of the lower mode, show the minimum energy $\omega_*$ in the region $d \gtrsim 2.36$ where the lowest mode is a magneto-roton.  The mass gap is shown by the curve, which is solid blue for $d \lesssim 2.36$ and dashed magenta for $d \gtrsim 2.36$.}
\label{fig:massvsd}
\end{figure}

\section{Numerical results}
\label{sec:numerics}

\subsection{Spectrum}
\label{sec:spectrum}

We compute the spectrum by finding the normalizable solutions at $k=0$.  The masses $\omega_0 \equiv \omega(k=0)$ of the lowest eight modes as a function of temperature for representative values of the other parameters are shown in Fig. \ref{fig:spectrum}.  As expected for a QH fluid, we see a mass gap; the size of the gap, shown Fig. \ref{fig:massvsd}, is given by the energy of the lowest mode.

The mass squared is positive for all modes, indicating perturbative stability, with one exception.  In \cite{Jokela:2011eb} we found in the vicinity of the high-temperature phase transition that there are two branches of MN solutions, and the MN embedding with smaller $r_0$ was found to be thermodynamically unstable.  The fluctuation analysis shows that this small-$r_0$ MN solution is perturbatively unstable, as well.  

\begin{figure}[ht]
\center
\includegraphics[width=0.7\textwidth]{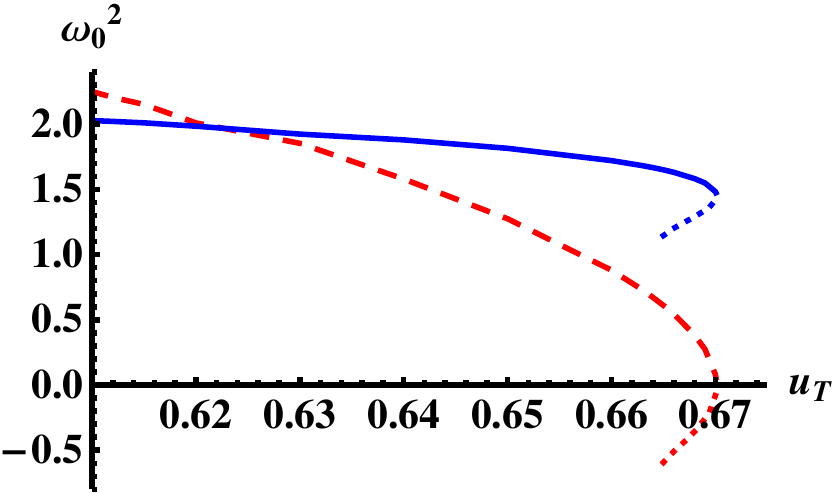}
\caption{The mass squared $\omega_0^2$ of the two lowest modes plotted against $u_T$, for $d=b=1$ and $m=0.1$.  The solid curves are the fluctuations of the stable MN solution; blue is a vector and red is a scalar.  The dashed lines are the fluctuations of the unstable solution.  The dashed red curve has $\omega_0^2 < 0$, with a positive imaginary part, and is therefore tachyonic.}
\label{fig:unstablebranch}
\end{figure}

Fig.~\ref{fig:unstablebranch} shows $\omega_0^2$ for the lowest two modes for both large-$r_0$ and small-$r_0$ branches as functions of temperature.\footnote{For ease of comparison, we have chosen the same parameters for Fig. \ref{fig:unstablebranch} as for Figs.~7 and 8 in \cite{Jokela:2011eb}.}  The lowest mode of the unstable branch (dashed red curve) is tachyonic.  The two branches meet at $u_T \approx 0.67$, just as the lowest mode of the stable solution (solid red curve) becomes massless.

Another feature we find at $k=0$ is a decoupling, also observed in \cite{Jokela:2010nu}, of the scalar $\delta\tilde\rho$ fluctuation and the vector ($\delta\tilde e_x$, $\delta\tilde a_y$) fluctuations.  At low temperature, the lowest mode is a vector, followed by a scalar.  Subsequently, every third mode is another scalar.  At nonzero $k$, the equations of motion are fully coupled, but we can continue to employ their $k=0$ labels.

From Fig.~\ref{fig:spectrum}, it is evident that all of the masses decrease with increasing temperature, though at different rates.   In particular, the masses of scalar modes decrease faster than the masses of vector modes, leading to periodic levels crossings.  As the temperature is raised, the crossing of the two lowest modes, shown in Figs.~\ref{fig:spectrum} and \ref{fig:unstablebranch}, occurs first.  At higher temperatures, the fourth and fifth modes cross and then the seventh and eighth modes, as shown in Fig.~\ref{fig:spectrum}.  While there is the potential for a crossing at every third mode, for a given choice of parameters, a finite number of crossings will occur before the maximum temperature is reached and the perturbatively stable branch of MN embeddings terminates.  

We have described the spectrum as a function of increasing temperature at certain fixed values of the charge density (and magnetic field).  However, the behavior is qualitatively the same if we hold the temperature fixed and decrease the density.

\subsection{Dispersion}
\label{sec:dispersion}
\begin{figure}[ht]
\center
\includegraphics[width=0.7\textwidth]{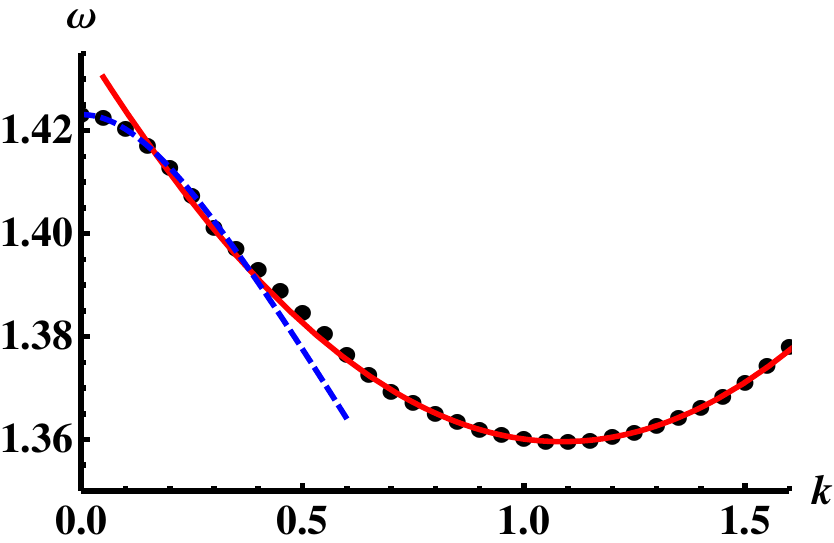}
\caption{The dispersion of the lowest mode at $d=1$, $u_T = 0.61$, $b=1$, and $m=0.1$.  The black dots are the numerical result, the solid red curve is the fit to the rotonic form (\ref{rotondispersion}) with $\omega_* = 1.360$ and $k_* = 1.0867$, and the dashed blue curve is the small-$k$ fit to (\ref{omega1}) with $\omega_0 = 1.461$ and $\delta = 0.0377$.}
\label{fig:dispersionfit}
\end{figure}

We now consider the dispersion of these modes for nozero $k$.  In most cases, the modes have massive dispersion relations, well-approximated by
\be
\label{massivedispersion}
\omega = \sqrt{\omega_0^2 + c_s^2k^2} \ ,
\ee
where $c_s$ is the speed of sound.  However, near the crossing of a scalar and vector mode, the lower of the two modes 
develops a rotonic dispersion, with a minimum $\omega_* < \omega_0$ located at $k_* > 0$, which is well approximated by the form
\be
\label{rotondispersion}
\omega = \sqrt{\omega_*^2 + c_s^2(k-k_*)^2} \ ,
\ee
as is shown in Fig.~\ref{fig:dispersionfit}.  However, near $k = 0$, the dispersion instead takes the form $\omega \sim \omega_0 - O(k^2)$.  This deviation is a result of the level crossing and will be addressed in Section \ref{sec:levelcrossingandrotons}. 

\begin{figure}[!ht]
\center
\includegraphics[width=0.7\textwidth]{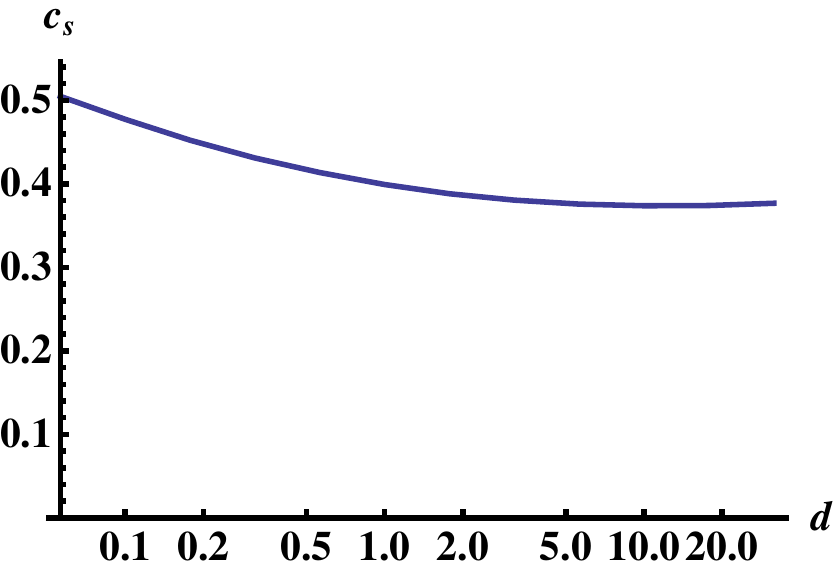}
\caption{The speed of sound $c_s$ for the lowest mode plotted against the charge density $d$ in a log scale for $u_T = 0.1$, $b=1$, and $m=0.1$.  Note that for small $d$, the mode is massive, while for $d \gtrsim 2.2$, it is rotonic.}
\label{fig:speedofsound}
\end{figure}

We compute the speed of sound by fitting the large $k$ numerical data to (\ref{massivedispersion}) and (\ref{rotondispersion}).  We find that $c_s$, defined by the large $k$ asymptotics of the dispersion, is fully
independent of the mode number and whether the mode is massive or rotonic.  The speed of sound is weakly dependent on the charge density; it increases slightly as $d$ is reduced, as shown in Fig.~\ref{fig:speedofsound}.

\begin{figure}[ht]
\center
\includegraphics[width=0.7\textwidth]{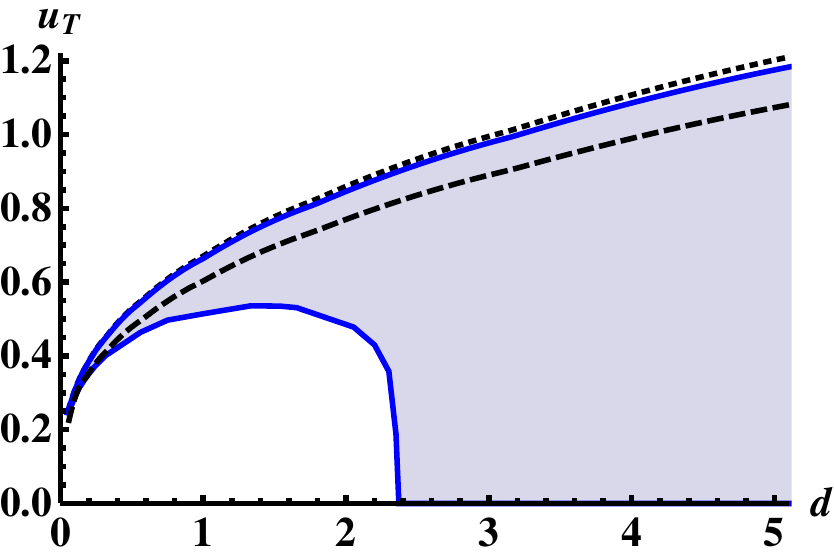}
\caption{The region in the $(d, u_T)$-plane in which the lowest mode is a magneto-roton, for $b=1$ and $m=0.1$, bounded by the blue solid curves.  The black dashed curve is the location of the first-order phase transition, and the dotted black curve is where the MN embedding solution ceases to exist.  Although the region near $d=0$ is numerically difficult to investigate, we have verified that the magneto-roton exists at least down to $d=0.1$, where the upper and lower bounds are at most $\delta u_T \approx 0.02$ apart.}
\label{fig:rotonregion}
\end{figure}

We will first focus on the magneto-roton appearing in the lowest mode; the magneto-rotons in higher modes are qualitatively similar.  We show in Fig.~\ref{fig:rotonregion} the region in the $(d, u_T)$ parameter space where this magneto-roton exists.  For small values of the charge, $d \lesssim 2.4$, the magneto-roton appears for a narrow range of temperatures, while for large charge, the magneto-roton exists all the way to zero temperature.  Although the upper bound on the magneto-roton range is quite close to the upper bound on the existence of the MN embedding, there is a small gap between the two. 

\begin{figure}[ht]
\center
\includegraphics[width=0.46\textwidth]{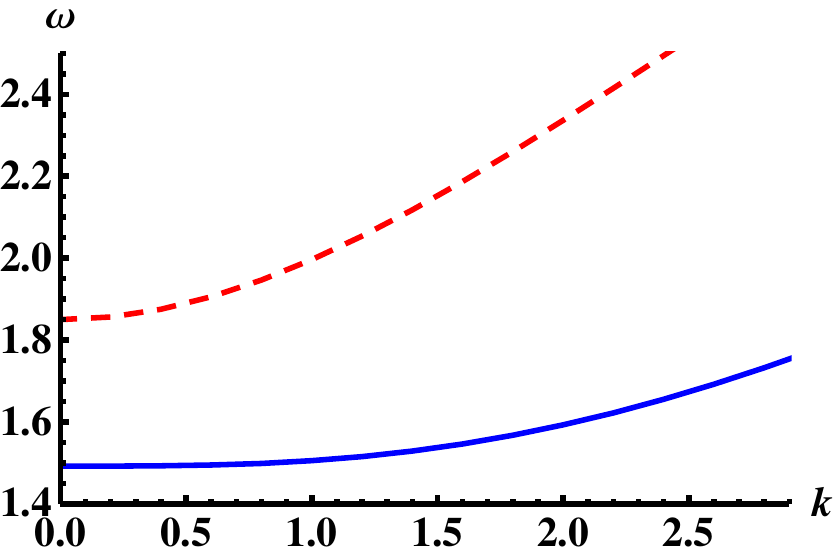}
\includegraphics[width=0.46\textwidth]{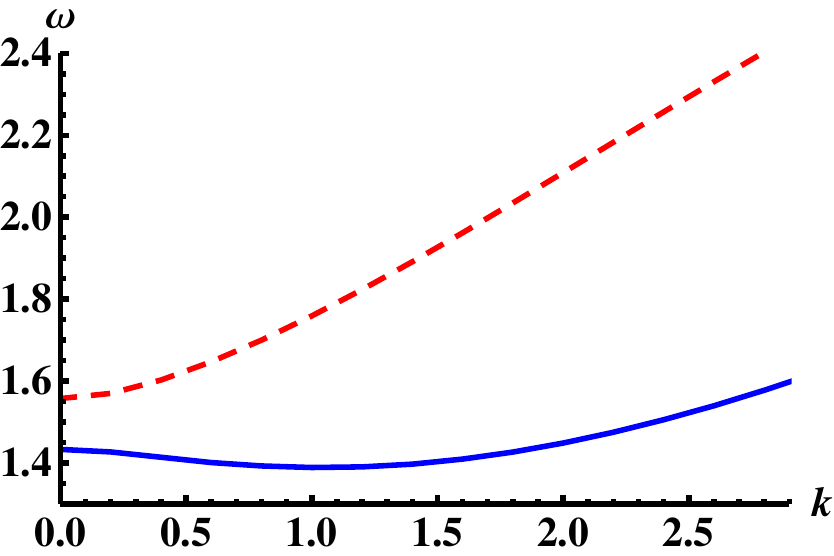}
\includegraphics[width=0.46\textwidth]{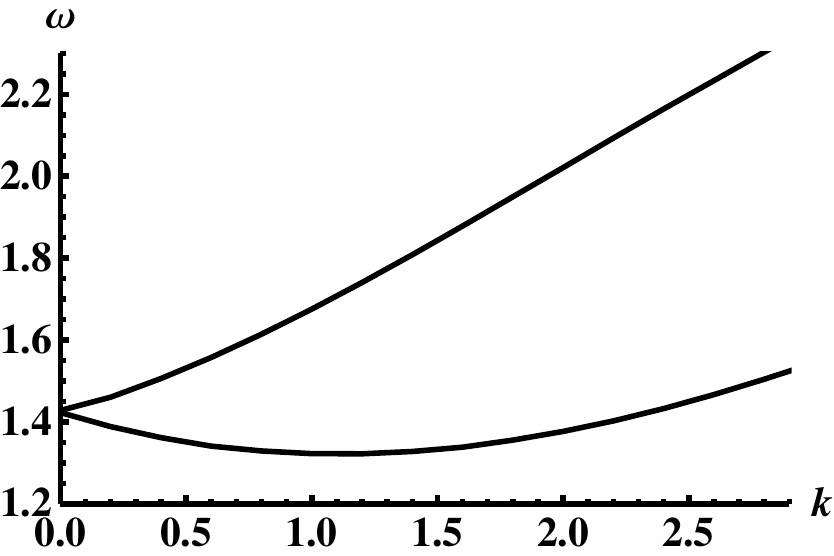}
\includegraphics[width=0.46\textwidth]{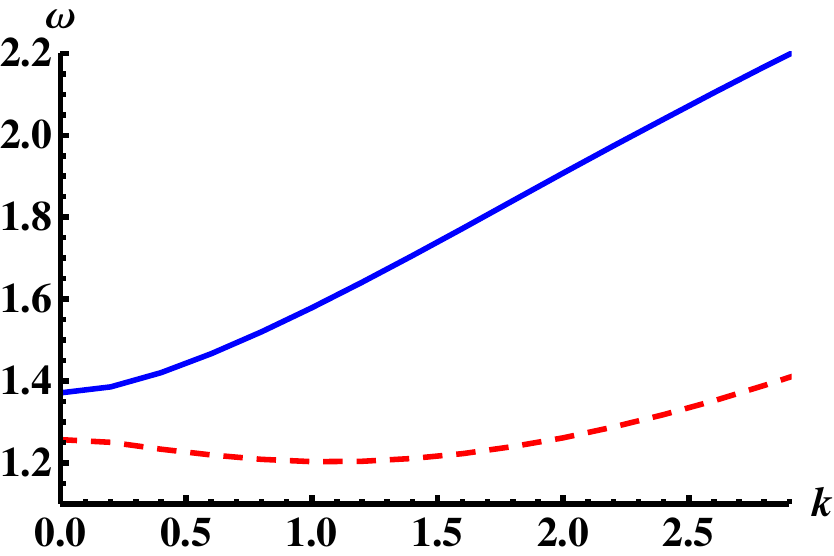}
\caption{The dispersions $\omega$ vs.~$k$ of the lowest two modes, the scalar in red and the vector in blue, are shown at four temperatures, $u_T = 0.51,0.60,0.62$, and $0.64$ with $d=b=1$ and $m=0.1$.  The upper left panel shows $u_T = 0.51$; the modes both have the standard massive dispersion.  In the upper right, where $u_T = 0.60$, the vector mode has become rotonic.  At $u_T = 0.62$, in the lower left, the two modes are degenerate at $k=0$, so the scalar and vector modes can not be distinguished; we denote this by plotting both in black.  In the lower right, at $u_T = 0.64$, the scalar is now a magneto-roton.}
\label{fig:rotondispersion}
\end{figure}

\begin{figure}[!ht]
\center
\includegraphics[width=0.46\textwidth]{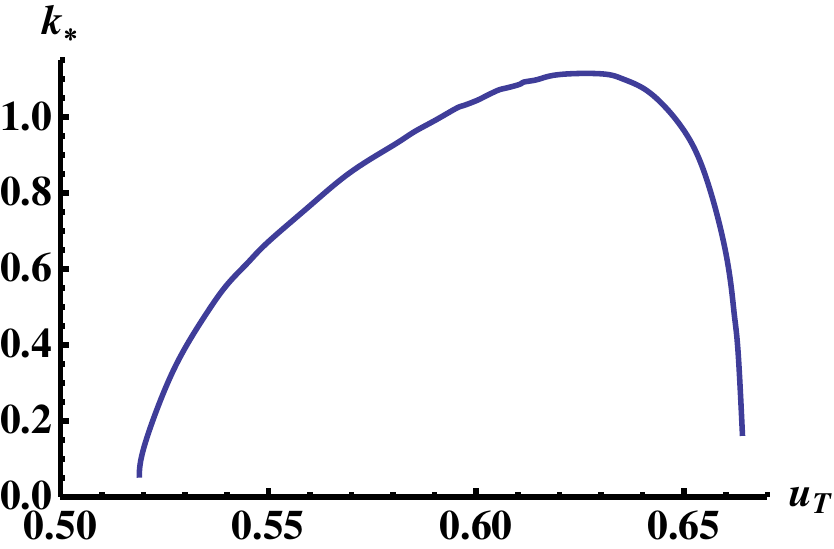}
\includegraphics[width=0.46\textwidth]{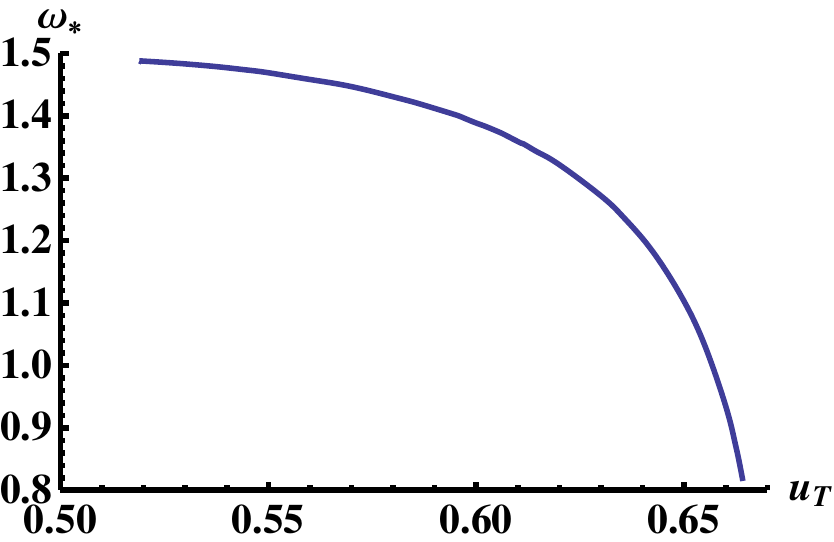}
\caption{The location $\tilde k_*$ of the magneto-roton minimum (left) and the minimum of the magneto-roton energy $\omega_*$ (right) plotted as functions of temperature $u_T$. Both figures are for $d=b=1$ and $m=0.1$.}
\label{fig:rotonlocation}
\end{figure}

Specializing for definiteness to $d=1$, Fig.~\ref{fig:rotondispersion} shows the dispersions of the lowest two modes for four different temperatures, and Fig.~\ref{fig:rotonlocation} plots the location and energy of the magneto-roton minimum, $k_*$ and $\omega_*$, as functions of the temperature.  This magneto-roton exists in a finite range, $0.52 \lesssim u_T \lesssim 0.665$ around the level crossing, which for these parameters occurs at $u_T\approx 0.62$.  Note that there is a small but nonzero distance between the upper bound and the maximum temperature, $u_T \approx 0.67$.  Right at the level crossing, the magneto-roton minimum reaches a maximum $k_*$.

\begin{figure}[th]
\center
\includegraphics[width=0.46\textwidth]{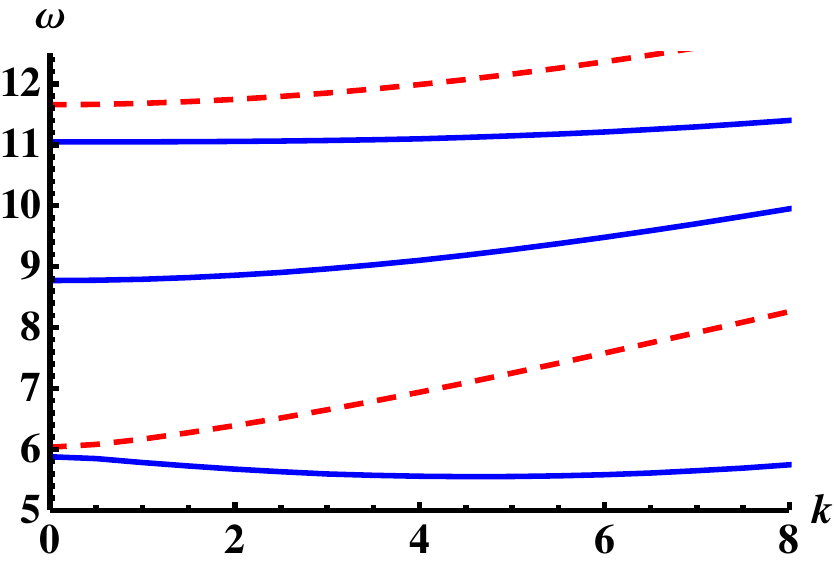}
\includegraphics[width=0.46\textwidth]{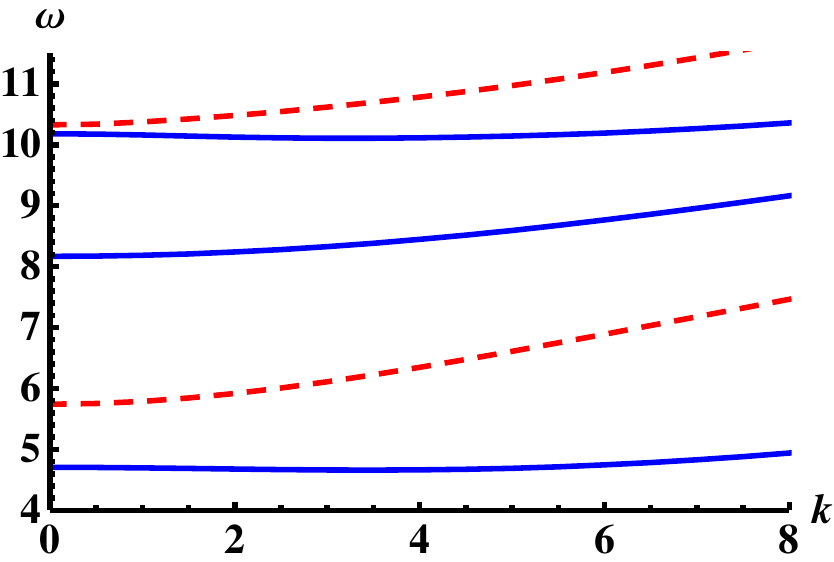}
\includegraphics[width=0.46\textwidth]{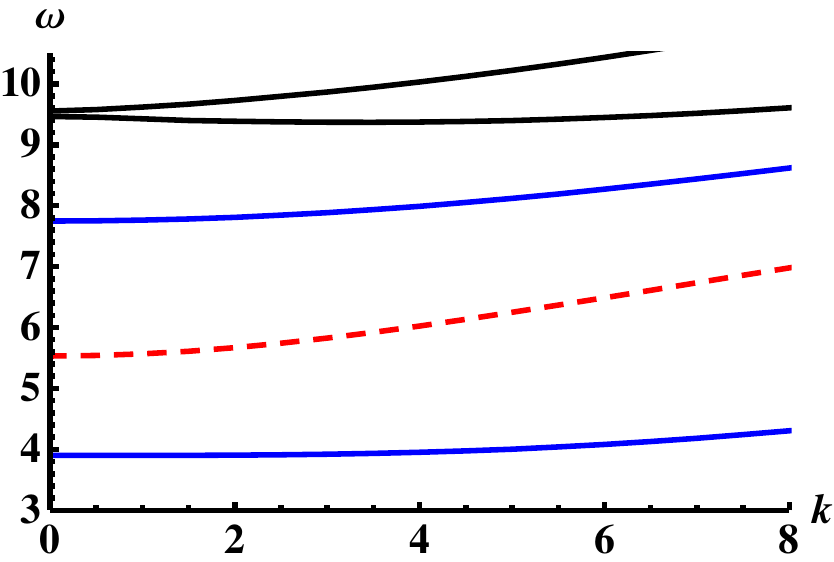}
\includegraphics[width=0.46\textwidth]{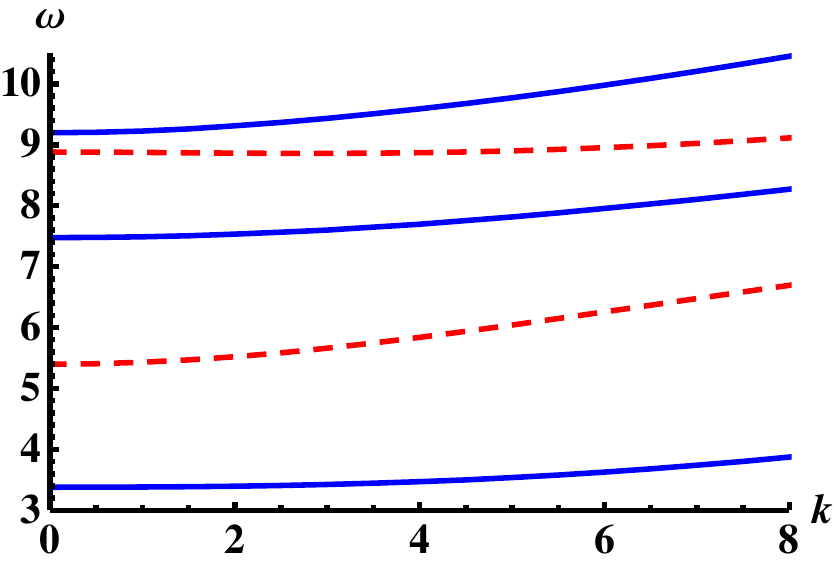}
\caption{The dispersions $\omega$ vs.~$k$ of the lowest five modes, scalars in blue and vectors in red, are shown at four temperatures, $u_T = 1.3, 1.45, 1.50,$ and $1.52$ with $d=10$, $b=1$ and $m=0.1$.  In the upper left panel, at $u_T = 1.3$ there is a magneto-roton only in the lowest mode.  In the upper right, where $u_T = 1.45$, the fourth mode, a vector, has become rotonic as well.  At $u_T=1.50$, in the lower left, the fourth and fifth modes are degenerate at $k=0$ and so are shown as black, with the fourth mode rotonic.  In the lower right, where $u_T= 1.52$, the scalar is now the fourth mode and is a magneto-roton.}
\label{fig:higherrotondispersion}
\end{figure}

The magneto-rotons observed in the fourth and higher modes are qualitatively much the same as the magneto-roton in the lowest mode.  They occur near level crossings between scalar and vector modes, although the regions of parameter space over which they exist are much smaller.  Fig.~\ref{fig:higherrotondispersion} is the analog of Fig.~\ref{fig:rotondispersion} for the fourth mode, showing the rotonic dispersion at several temperatures.  As is evident from Fig.~\ref{fig:spectrum} and Fig.~\ref{fig:higherrotondispersion}, the crossings of higher levels, and thus the higher magneto-rotons, occur at higher temperatures.  As the temperature is increased, we generally find a narrow region where two magneto-rotons coexist.

\section{Level crossing and magneto-rotons}
\label{sec:levelcrossingandrotons}

Let us perform a brief analytical analysis to clarify the relationship between the level crossings and consequent rotonic dispersion relations. The solutions to the fluctuation equations of motion define a linear function which maps the IR boundary conditions to the UV ones. We may define its $3\times 3$ matrix representation as
\be
 \left(\begin{array}{c} \delta \tilde e_x(\infty) \\ \delta \tilde a_y(\infty) \\ \delta \tilde \rho (\infty) \end{array}\right) = M\left(\begin{array}{c} \delta \tilde e_x(0) \\ \delta \tilde a_y(0) \\ \delta \tilde \rho (0) \end{array}\right) \ . 
 \ee
The matrix depends on the external parameters $u_T$ and $d$ as well as on $\omega$ and $k$.

As pointed out above, for $k=0$ the off-diagonal elements of $M$ mixing the scalar $\delta \tilde \rho$ and vector $(\delta \tilde e_x, \delta \tilde a_y)$ fluctuations vanish. At $k=0$, we can diagonalize $M$ by choosing an appropriate basis for the vector fluctuations,
\be
 \delta e_\pm = \alpha \, \delta \tilde e_x \pm i \beta \, \delta \tilde a_y \ ,
\ee
where $\alpha$ and $\beta$ are real coefficients which must be determined numerically. Notice that this 
change of basis is not a unitary transformation. 

Consider further a choice of $u_T$ and $d$ such that a scalar mode and a vector mode have almost equal energies, say $\omega_\pm \equiv\omega_0 \pm \delta$, where $\delta/\omega_0 \ll 1$, and the plus sign corresponds to the vector mode. Let us fix the coefficients $\alpha$ and $\beta$ such that $M$ is diagonal at $\omega=\omega_+$ and $k=0$, with $\delta e_-$ corresponding to the normalizable fluctuation mode. We order the new basis vectors as $(\delta e_+,\delta e_-,\delta \tilde \rho)$. Then only $M_{11}$ and $M_{33}$ are nonzero in the new basis when evaluated at the point defined above.\footnote{We could also let $\alpha$ and $\beta$ depend on $\omega$ such that at $k=0$, $M$ would be diagonal for all $\omega$.}  Expanding to the first order in $\delta$ and $k$ around this point gives
\be
 M \approx  \left(\begin{array}{ccc} a_{11} (\omega-\omega_+) +c_{11} & a_{12}(\omega-\omega_+) & 0 \\
  -a_{12}(\omega-\omega_+) & a_{22}(\omega-\omega_+) & 0\\
 0 & 0 & a_{33}(\omega-\omega_-) \end{array}\right) + \left(\begin{array}{ccc} 0 & 0 & b_{13} \\
    0 & 0 & b_{23}\\ b_{31} & b_{32} & 0 \end{array}\right) k \ .
\ee
Notice that the linear $k$-dependence only appears in the mixing between the scalar and vector modes.  This reflects a discrete symmetry of the fluctuation equations of motion.  Indeed we can check that the Lagrangian densities of~\eqref{DBIfluctuations} and~\eqref{CSfluctuations} are invariant up to $\mathcal{O}(\epsilon^2)$ (but not at higher orders) under the transformation 
\be
 x \mapsto -x \ ; \qquad y \mapsto -y \ ; \qquad \delta a_{x,y}  \mapsto -\delta a_{x,y} 
\ee
which leads to the symmetry
\be
 k \mapsto -k \ ; \qquad \delta \tilde a_{y}  \mapsto -\delta \tilde a_{y}  \ ; \qquad \delta \tilde e_{x}  \mapsto -\delta \tilde e_{x} 
\ee
of the equations of motion for the fluctuation.  Also the matrix $M$ must be covariant under this transformation, which explains its structure.
 
For normalizable modes we require, as above, 
\be 
\det M = 0 \ .
\ee
Calculating the energies of these modes from the approximate matrix, to first order in $\delta$ and $k$, gives
\bea 
\label{omega1}
\omega_1 &=& \omega_0 - \sqrt{\delta^2+\frac{b_{23} b_{32}}{a_{22} a_{33}}k^2} \\
\label{omega2}
\omega_2 &=& \omega_0 + \sqrt{\delta^2+\frac{b_{23} b_{32}}{a_{22} a_{33}}k^2} \\
 \omega_3 &=& \omega_+ - \frac{c_{11} a_{22} }{a_{11} a_{22}+ a_{12}^2} \ .
\eea
The dispersions (\ref{omega1}) and (\ref{omega2}), with the values of $a_{ij}$ and $b_{ij}$ extracted from numerics,  give the small-$k$ behavior of modes near a level crossing and match the behavior observed in Figs.~\ref{fig:dispersionfit}, \ref{fig:rotondispersion}, and \ref{fig:higherrotondispersion}.  In particular, the lower mode, $\omega_1$, initially decreases with $k$.  On the other hand, at large $k$, the energy will instead be increasing like $c_s k$; there must necessarily be a local minimum at some nonzero $k$.  This magneto-roton minimum is therefore a direct consequence of the level crossing.

Another feature of (\ref{omega1}) and (\ref{omega2}) which we find in Figs.~\ref{fig:rotondispersion} and \ref{fig:higherrotondispersion} is the change in $k$-dependence at the point where the modes are degenerate.  When $\delta > 0$, the first $k$-dependent correction to the energy is quadratic: $\omega \sim \omega_\pm + O(k^2)$.  However, when the modes are degenerate and $\delta = 0$, the dependence on the momentum is linear $\omega \sim \omega_0 + O(k)$ .  The dependence on the expansion coefficients appears only trough the combination $b_{23} b_{32}/(a_{22} a_{33})$, which is invariant under arbitrary scalings of the basis vectors $\delta e_-$ and $\delta \tilde \rho$.  In order for the energies to remain real as $k$ increases, this combination must be positive. While it is difficult to exclude the apparently spurious case of negative $b_{23} b_{32}/(a_{22} a_{33})$ by analytical methods, 
it did not show up in any of the level crossings that we studied numerically.
Notice also that the result depends only on the mixing coefficients $b_{32}$ and $b_{23}$ between the two nearby modes. In fact, ignoring the other mode and simply diagonalizing the lower-right $2 \times 2$ subblock of $M$ gives exactly (\ref{omega1}) and (\ref{omega2}).

The energy of the additional vector mode $\omega_3$ does not depend on $k$ to linear order, as is expected for modes in the absence of level crossing. Since this mode does not lie within $\sim \delta$ of the modes which are about to cross, the linear expansion gives at best a rough approximation to its energy.  However, we may improve the precision of $\omega_3$ by adding higher order corrections in $\omega$ in the calculation, which does not change the $k$-dependence of $M$ as it is protected by the discrete $k \to -k$ symmetry.  Therefore, linear terms in $k$ will continue to be absent in the expression for $\omega_3$.


\section{Discussion}
\label{sec:discussion}

We continued in this paper our investigation of the D2-D8' model by analyzing the fluctuations of the QH state at nonzero temperature.  We found that the neutral spectrum is stable and gapped, providing evidence that the MN embedding describes an incompressible QH fluid.  We also showed that the thermodynamically unstable phase is perturbatively unstable, as well. 

As observed in real QH fluids, the lowest mode was found to have a rotonic dispersion, at least for much of the parameter space.  In addition, we uncovered magneto-rotons in higher modes as well.  We showed that all of these magneto-rotons are direct consequences of near degeneracies at zero momentum between a scalar mode and a vector mode.  Experiments have detected two such closely spaced, long-wavelength modes \cite{pinchuz}.  In  \cite{tokatly}, a theoretical treatment has been given in terms of a hydrodynamical model.  A composite fermion model has also been proposed \cite{jain1, jain2} and identifies the second mode as a two-magneto-roton bound state.

Many of our results for the D2-D8' model are qualitatively similar to those found in \cite{Jokela:2010nu} for the D3-D7' system.  While both models found magneto-rotons in the lowest mode, in the D3-D7' model a magneto-roton was only found in a narrow region in parameter space and no magneto-rotons were seen in higher modes.  However, D3-D7' investigation was restricted to zero temperature, and it potentially could be that magneto-rotons are more common at higher temperatures.

Of course, these types of $\# ND=6$  brane constructions are far from perfect holographic models of QH systems, which is not surprising given that they contain in their construction many unphysical features, such as large $N$.
Furthermore, there is a parametrically large hierarchy between the mass gap for charged excitations and for neutral excitations, $\omega_0^{charged}/\omega_0^{neutral} \sim \lambda^{2/5}$, which is absent in real quantum Hall systems.  However, the quantum Hall effect is a very robust phenomenon in (2+1)-dimensional fermion systems, so we are hopeful that there is a large universality class of models which exhibit its characteristic properties.  The close qualitative similarity between the D2-D8' and D3-D7' systems, we believe, suggests that these types of brane models lie within such a universality class.

We restricted our attention in this paper to the normal modes of the MN embeddings of the D8-brane; in a future work, we will address the properties of the quasi-normal modes of the BH embeddings, which holographically model a Fermi-like liquid.  The analogous quasi-normal mode analysis of the D3-D7' model was performed in \cite{Bergman:2011rf}.  The neutral spectrum was found to be ungapped and to feature a zero-sound mode.  Interestingly, at sufficiently high charge density, a tachyon appears at nonzero momentum, indicating  an instability to a spatially-modulated, spin and charge density wave state.  As both brane models contain similar Chern-Simons terms of the type which generate this sort of instability, we anticipate that it should appear in the D2-D8' system, as well.

\bigskip
\noindent

{\bf \large Acknowledgments}

We would like to  thank Oren Bergman,  Gilad Lifschytz, John McGreevy, and Andrei Parnachev for helpful discussions, comments, and suggestions.  N.J. is supported in part by the Israel Science Foundation under grant no.~392/09 and in part at the Technion by a fellowship from the Lady Davis Foundation.  N.J. wishes to thank the University of Santiago de Compostela for hospitality while this work was in progress.  N.J. also wishes to thank the Institute for Nuclear Theory at the University of Washington for its hospitality during the completion of this work.  M.J. is supported in part by Regional Potential program of the E.U.FP7-REGPOT-2008-1: CreteHEPCosmo-228644 and by Marie Curie contract PIRG06-GA-2009-256487.  The research of M.L. is supported by the European Union grant FP7-REGPOT-2008-1-CreteHEPCosmo-228644.  M.L. would also like to thank both the Technion and the Lorentz Center at the University of Leiden for their gracious hospitality.

\appendix

\end{document}